\documentclass[twocolumn,preprintnumbers,amsmath,amssymb,prl,showkeys,showpacs]{revtex4}

\makeatletter
\renewcommand{\section}{\@startsection{section}{1}{0mm}{-2\baselineskip}
{-0.5\baselineskip}{\normalfont\normalsize\bf}}
\renewcommand{\subsection}{\@startsection{subsection}{2}{0mm}{-\baselineskip}
{-0.5\baselineskip}{\normalfont\normalsize\bf}}
\newcommand{\d@t}{.}
\makeatother

\newcommand{\nn}{\nonumber}
\newcommand{\p}{\partial}

\newcommand{\Tr}{\mbox{Tr}}

\def\bea{\begin{eqnarray}}
\def\eea{\end{eqnarray}}
\def\be{\begin{equation}}
\def\ee{\end{equation}}

\begin{document}

\title{Comments on the topological open membrane}


\author{B. Pioline}
\email{pioline@lpthe.jussieu.fr}
\affiliation{LPTHE, Universit\'es Paris VI et VII, 4 pl Jussieu, \\
75252 Paris cedex 05, France}


\preprint{LPTHE-02-01, {\tt hep-th/0201257}}

\begin{abstract}Just as non-commutative gauge theories arise
from quantising open strings in a large magnetic field, non-Abelian
two-form gauge theories may conceivably be constructed by
quantising open membranes in a large three-form magnetic background. We
make some observations that arise in following this strategy,
with an emphasis on the relation to the quantisation 
of volume-preserving diffeomorphisms (vpd). In particular,
we construct consistent 
non-Abelian interactions of a two-form in 3+1 dimensions,
based on gauge invariance under vpd.
\end{abstract}



\maketitle

Of the many mysteries surrounding the theory formerly known as string theory,
the dynamics of the five-brane is perhaps the most intriguing. 
While a single type IIA or M5
brane supports a free self-dual two-form gauge field on its world-volume,
together with five transverse scalar degrees of freedom and their fermionic
partners \cite{Callan:1991ky}, the case of 
$N$ five-branes at small separation is much less
understood. Analogy with the case of coinciding D-branes and duality
with ALE singularities in type IIB string theory suggest that the
chiral multiplet now transforms in the adjoint representation of a
spontaneously broken $U(N)$, with the open membranes stretched between two
five-branes playing the r\^ole of stringy $W$ bosons on the five-brane
worldvolume \cite{Witten:1995ex}. This picture is 
at best suggestive however, since the quantization of the membrane,
if at all sensible, is still beyond reach
(see \cite{Dasgupta:2002iy} for a review of recent attempts however). 
Disregarding the fundamental
origin of these low-energy degrees of freedom, no consistent local theory
of interacting non-Abelian self-dual two-forms has been found to 
date, in line with no go theorems \cite{Bekaert:1999dp}. 
The nature of the $U(N)$ non-Abelian symmetry itself is unclear, since anomaly
and entropy considerations hint at $N^3$ degrees of freedom rather than
$N^2$ as expected on a naive perturbative basis \cite{Klebanov:1996un} --
an admittedly unwarranted expectation due to the absence of a tunable
coupling.

On the other hand, recent progress in the understanding of D-branes in
background fields has taught us that non-Abelian dynamics arise even
at $N=1$, in the presence of a strong background magnetic or electric 
field \cite{Connes:1997cr}.
Indeed, the effects of the higher-derivative Born-Infeld couplings can be
resummed by going to a non-commutative description, where the $U(1)$
gauge invariance $\delta a_\mu = \p_\mu\lambda$ is replaced by
a transformation $\delta a_\mu = \p_\mu\lambda + a_\mu*\lambda-\lambda*a_\mu$
formally identical to the usual Yang-Mills gauge 
invariance \cite{Seiberg:1999vs}.
The Moyal deformation of the ordinary commutative product can in fact
be derived without detailed knowledge of the dynamics of open strings
in a magnetic background $B=dA$: in the limit of large $B$, one may retain
only the topological coupling $\int_{\Sigma} B$ from the action of the 
string with worldsheet $\Sigma$ (we omit the pull-back from target space).
This in turn reduces to a coupling $\oint_\gamma A$ on the boundary,
hence to a topological quantum mechanical model. For constant background,
the action $\oint B_{ij} X^i(\tau) [d X^j/d\tau] d\tau$ is Gaussian and
first order in time derivatives, and yields the Moyal product structure
for the correlators $\langle \prod_{i=1\dots p} f_i(X(\tau_i))\rangle
=\int d^n x f_1 * \dots * f_p (x)$. To leading order, the effect of
the strong magnetic field is therefore to dress the free Maxwell
action with Moyal products \cite{Seiberg:1999vs}.

From this point of view, it is very tempting to try and apply the same
logic to the M5-brane. In the presence of a strong magnetic three-form 
field strength $H$, the self-duality equations become highly non-linear
\cite{Aganagic:1997zq}, 
and it is conceivable that these higher-derivative effects may 
be resummed in terms of a non-commutative deformation of the linear
self-duality equations~\footnote{Due to the self-duality 
condition we avoid mentioning the action and focus on the equations
of motion only.}, hopefully provided by the quantization
of the topological open membrane. If one can furthermore find a matrix 
realization formally isomorphic
to this non-commutative deformation, one will have succeeded in producing
a non-Abelian extension -- hopefully the only one -- of the free tensor 
dynamics. In this note, we present some observations that arose in 
following this strategy, in the hope that they will serve in attaining
this goal. Our attempt at quantizing
the topological open membrane is not by any means the first, see e.g.
\cite{Bergshoeff:2000jn,Kawamoto:2000zt,Park:2000au,Matsuo:2001fh,Das:2001mg},
however we hope to provide a different perspective 
on this still unsolved problem. 
Our salient result is a consistent deformation of the 
dynamics of an Abelian two-form in 3+1 dimensions, which uses the
group of volume-preserving diffeomorphisms in 3 dimensions as a gauge group,
and which will be presented in 
the last section. The main message we want to convey is that
the non-Abelian dynamics of tensor multiplets and the quantization of the 
open membrane are closely connected to the quantization of the 
volume-preserving diffeomorphisms in 3 dimensions, or rather to its
realisation by means of the Nambu bracket. 

\section{Topological open membranes.}
By analogy with the string, let us consider the dynamics of membranes
in the background of a strong three-form field $H$. The coupling 
of a supermembrane to the background field occurs through the topological 
``Wess-Zumino'' term $\int H$ only, which for 
closed membranes simply yields a phase factor
proportional to the flux of $H$ through the 3-cycle \cite{Bergshoeff:1987cm}. 
This flux vanishes
in the absence of a non-trivial topology in target space. Open membranes 
are charged under $H$ -- however they only exist in the presence of
five-branes on which they can end \cite{Witten:1995ex,Ezawa:1997xc}
~\footnote{Membranes could also end on
Horava-Witten 9-branes, or on one-dimensional 
defects (see e.g. \cite{Dasgupta:2002iy}).}. 
The field $H$ is therefore the self-dual
field strength of the two-form gauge field living on the five-brane 
worldvolume, plus the self-dual part of the bulk three-form gauge field
$C$ for gauge invariance. Due to the non-linear self-duality constraint
on $H$, one may worry whether a large $H$ limit exists. As discussed in
\cite{Bergshoeff:2000jn}, a generic constant non-linear 
self-dual three form $H$ may be written
in a particular Lorentz frame as (a) $H=\tanh\phi~ dx^{045}
+\sinh\phi~ dx^{123}$, so that there is a limit 
in which the magnetic component $H_{123}$ becomes very
large while the electric component $H_{045}$ saturates
to its critical value \cite{Gopakumar:2000ep}. Even though the 
motions in the planes $123$ and $045$ do not decouple
due to the membrane non-linearities, it may be useful to understand
the effects of the large $H_{123}$ and of the critical $H_{045}$ fields
separately. There also exists a non-generic ``light-like''
case (b) $H=e^{\phi} (dx^0-dx^1)(dx^{23}+dx^{45})$, where all components
become large as $\phi\to\infty$, although $H^2=0$ 
throughout \cite{Aharony:1997an}. With these preliminaries in mind, 
the action for an open membrane in the strong $H$ limit reduces to,
after integrating by part, 
\begin{equation}
\label{topact}
S=\int_\Sigma H_{ijk} ~ X^i dX^j \wedge dX^k
\end{equation}
where $\Sigma$ is the membrane boundary, which we shall take connected
for simplicity, and $X^{i}$ are the embedding coordinates of the
membrane, fonctions of the worldsheet coordinates $(\sigma,\tau)$
($X^{1,2,3}$ for case (a), or $X^0-X^1,X^{2,3,4,5}$ in case (b),
since the coordinate $X^0+X^1$ decouples). As noticed in 
\cite{Matsuo:2001fh}, this is
also the action for a vortex line in an incompressible and
inviscid 3D fluid such as superfluid He II (after adding in the
kinetic energy of the fluid, $\int v^2 d^3x$ where $v$ is the
velocity field of the fluid with vorticity localized along the 
line) \cite{Lund:1976ze}.
This action is symmetric under general diffeomorphisms on the worldsheet
$\sigma$, and under diffeomorphisms preserving the three-form $H$ in target
space -- we'll refer to the latter as to volume-preserving diffeomorphisms 
(vpd), or self-dual volume-preserving diffeomorphisms (svpd)
for case (b). Our hope is that despite the strong non-linearity 
of \eqref{topact}, these symmetries will allow to solve for the 
quantum theory associated to it. This hope is perhaps not
unwarranted, in light of the existence of toy models with a cubic action
which are nonetheless free \cite{kazhdan} (see \cite{waldron} for an 
application of these models to the quantum BPS membrane). The theory
\eqref{topact} has already been considered from a canonical 
\cite{Bergshoeff:2000jn,Kawamoto:2000zt} or Batalin-Vilkovilsky 
\cite{Park:2000au}~\footnote{It may be useful to develop a BV quantization
of the two-dimensional theory \eqref{topact} itself, instead of
the three-dimensional theory with boundaries considered in \cite{Park:2000au}.
For that purpose, the equation $*dX^i=H_{ijk} X^j dX^k$ is a
very natural gauge-fixing function, and it would be interesting to study
the field configurations onto which it localizes.} point of view, 
however our approach will be rather
different, and, as in \cite{Matsuo:2001fh}, will emphasize the 
symmetry under vpd.

\section{Classical dynamics.}
The equations of motion $H_{ijk}~ dX^j \wedge dX^k=0$ following
from \eqref{topact} imply that the 
embedding coordinates $X^i$  are fonctions
of a single combination $f(\sigma,\tau)$ of the worldsheet coordinates.
Up to reparameterisation of the worldsheet, one may choose $f=\sigma-\tau$,
which shows that we are dealing with a chiral theory. The correlators 
of vertex operators such as $e^{i kX}(\sigma,\tau)$ will therefore be 
holomorphic in $z=\sigma-\tau$, with possible monodromies in the $z$ plane.
More conveniently, we may choose $f=\sigma$, and think of $\tau$ as 
the time coordinate: in the absence 
of sources, the solutions of the equations of
motion are therefore static loops $\gamma$ of arbitrary shape in target space. 
This is analogous to the statement that charged particles 
in two dimensions are frozen at a point in the presence of a 
strong magnetic field. The staticity is also a consequence of the
invariance by time reparameterization, which implies a vanishing
Hamiltonian. The canonical treatment of this system in case (a)
was carried out long ago in the context of vortex lines in superfluids
\cite{Rasetti:hk}, and rediscovered many times since then.
Since the time derivative appears in 
first order, the phase space is restricted by the primary constraints
$C_i=\pi_i-H_{ijk} X^j \p_\sigma X^k=0$ ~\footnote{Of course, these
constraints would be relaxed by keeping the kinetic term in the
action. The stringy choice $\int d\sigma d\tau (\p_\tau X^i)^2 -
(\p_\sigma X^i)^2$ made by Regge and Lund \cite{Lund:1976ze} does not seem
adequate for the membrane case at hand, however it does lead
to interesting soliton solutions \cite{zee}.}. The total momentum 
$\Pi_i=\int d\sigma~\pi_i=H_{ijk} \int_S dX^j dX^k$ 
is therefore equal to the flux of
$H$ through the surface $S$ bounded by the loop in target space. 
As in the case of dipoles in a magnetic field, the size of the
membrane increases proportional to its linear momentum. 
Of the three constraints $C_i$, the two orthogonal
to $\p_\sigma X^i$ are second class, while $C=C_i \p_\sigma X^i$
is first class, being the generator of the spatial diffeomorphisms.
One may treat all constraints as second class (i.e. restricting
the phase space of the system) by introducing a fourth
constraint $C_0=0$ fixing the spatial diffeomorphisms, such as 
$C_0=X^3-g(\sigma)$ ~\footnote{Another interesting choice is
$C_0=\sum (X^i)^2 -g(\sigma)$, which yields the commutation rules
for $SU(2)$ currents at level 0 (S. Minwalla, private communication).}. 
Neglecting global issues, one may set 
$g(\sigma)=\sigma$, and obtain
the canonical Dirac brackets 
\begin{equation}
\label{x1x2}
[X^1(\sigma),X^2(\sigma')]=\delta(\sigma-\sigma') \ .
\end{equation}
The embedding coordinates are therefore pointwise non-commutative,
hence the name ``non-commutative string'' given to the
topological open membrane by the authors of 
\cite{Bergshoeff:2000jn,Kawamoto:2000zt,Das:2001mg}. One may 
however legitimately worry that
all the interesting part of the dynamics lies in the global issues.
Alternatively, one may consider reparameterisation invariant observables only. 
Those take the form $J(A)=\int_\gamma A_i [dX^i/d\sigma] d\sigma$ 
where $A_i$ is a one-form gauge field in target space. Upon coupling
to $J(A)$ at time $\tau=0$, the equations of motion become
$H_{ijk} \p_\sigma X^j \p_\tau X^k = F_{ij} \p_\sigma X^j \delta(\tau)$
where $F_{ij}=\p_i A_j-\p_j A_i$ is the curvature of $A$.
For $d=3$, where $H_{ijk}=H \epsilon_{ijk}$, 
this implies that $[X^i]^{0^+}_{0^-}=\frac12
\epsilon^{ijk} F_{jk}/H$, meaning that the
loop has moved by the action of the diffeomorphism generated by
the vector $\xi=*dA/H$. It is essential to recognize that this
diffeomorphism is divergenceless, $d*\xi=0$, i.e.
it preserves the volume form $H$. Indeed the Dirac bracket
of the operators $J(A)$ may be readily computed, and reads
\begin{equation}
\left[ J(A) , J(A') \right] = J( A'')
\end{equation}
where $A''$ is the vector potential associated to
the volume-preserving diffeomorphism generated by the Lie bracket $[\xi,\xi']$
\cite{Rasetti:hk}. The open topological membrane therefore 
classically furnishes a representation of the algebra of volume-preserving
diffeomorphisms.

\section{Quantisation.}
In order to understand the structure of the Hilbert space of the
topological membrane, let us go back to the case of a charged 
particle in the limit of a strong magnetic field. The classical
trajectories consist of Larmor orbits of vanishing radius 
as the magnetic field goes to infinity. Quantum mechanically,
only the lowest Landau level remains, corresponding to a
Gaussian wave packet around the origin, with infinite degeneracy
(in unbounded geometry) arising from the action of area-preserving
diffeomorphisms moving the origin around.
In the present case, the classical configurations are now static
loops, which can be moved and deformed from one into another
under the action of the volume-preserving diffeomorphisms, 
so long as the topological invariants
of the loop are preserved. One may thus foresee an Hilbert space
consisting of all topological types of knots, with a degeneracy
generated by the group of vpd, together with harmonic degrees
of freedom around each loop, of infinite energy in the limit
under consideration. The quantisation of the topological membrane thus amounts
to the quantisation of the group of vpd (this point of view is also
stressed in \cite{Matsuo:2001fh}). Unfortunately, the quantisation 
based on the canonical commutation rules \eqref{x1x2} leads to singularities
in the correlators $\langle \prod_i J(\zeta_i e^{ik_i X^i}) \rangle$
whose proper treatment is rather unclear. In addition, the quantisation
of the group of vpd {\it stricto sensu} is known not to exist: the
cohomology group $H_2(G,G)$ classifying deformations at leading
order is non-trivial, but the corresponding deformation is obstructed 
at second order \cite{roger}. It is conceivable however that the structure
to be deformed is not precisely the group of vpd, but a structure
equivalent to it in the classical regime, as we now discuss.

\section{Vpd and Nambu dynamics.}
We have seen previously that a volume-preserving diffeomorphism
in 3 dimensions can be written as the divergence of a one-form,
$\xi=*dA$. By a slight extension of the Darboux theorem, it is
also possible to locally represent the two-form $*\xi$ as a
product $dH \wedge dK$, where $(H,K)$ are
a pair of symplectic local coordinates. The diffeomorphism
$\xi$ then acts on fonctions on the three-dimensional ``phase space''
$(x^1,x^2,x^3)$ through the Nambu bracket \cite{Nambu:1973qe},
\begin{equation}
\label{nambubra}
\delta_\xi F = \{ H, K, F \} :=
\epsilon^{ijk} \p_i H \p_j K \p_k F
\end{equation}
analogous to the Poisson bracket for area-preserving diffeomorphisms.
This bracket is a completely antisymmetric linear fonction of its
arguments, and satisfies the derivation property
\begin{equation}
\{f_1 f_2,f_3,f_4\}=f_1 \{f_2, f_3, f_4\} + \{f_1, f_3, f_4\} f_2
\end{equation}
and the so called Fondamental Identity \cite{Takhtajan:1994vr}
\begin{eqnarray}
\label{fi}
\{\{f_1,f_2,f_3\},f_4,f_5\} &+&
\{f_3,\{f_1,f_2,f_4\},f_5\} \\
+\{f_3,f_4,\{f_1,f_2,f_5\}\}&=&\{f_1,f_2,\{f_3,f_4,f_5\}\} \nonumber
\end{eqnarray}
generalizing the Jacobi identity of the Poisson bracket.
As is well known, this structure was introduced by Nambu as 
an alternative way to construct dynamical systems satisfying
the Liouville property, ie the conservation of probability
on phase space \cite{Nambu:1973qe} (see \cite{Nambu:1980kz} for an application
to the quantization of strings). Such systems satisfy the evolution equation 
\begin{equation}
\label{nambudyn}
dF/d\tau=\{H,K,F\}
\end{equation} 
for a pair of fixed generalized Hamiltonians $(H,K)$ 
and $F$ a function on phase space. 
Interestingly, these
equations of motion follow from an action 
functional\cite{Takhtajan:1994vr}~\footnote{The Euler-Lagrange
equations following from this action are $\epsilon_{ijk}
\left( {\partial X^j}/{\partial \tau}
-\{H, K, X^j\} \right)  {\partial X^k}/{\partial \sigma}=0$,
which imply \eqref{nambudyn} for $F=X^i$ after using
the symmetry under spatial diffeomorphisms.}
\begin{equation}
S=\int d\sigma~d\tau \left( \epsilon_{ijk} X^i \p_\sigma X^j \p_\tau X^k
- H dK d\tau \right)\ ,
\end{equation}
analogous to the action $S=\int pdq-Hdt$ in Hamiltonian dynamics,
which in the case of vanishing generalized Hamiltonians $H=K=0$ 
is precisely our topological membrane action \eqref{topact}. The
quantisation of the topological membrane is therefore equivalent
to the quantisation of Nambu dynamics. In particular, one would
like to find an analogue of geometric quantisation in which the
Nambu bracket is deformed while preserving its fundamental properties -
due to the ambiguity in the choice of Hamiltonians $(H,K)$ for a given
diffeomorphism $\xi$, this question is not strictly equivalent to
the quantisation of vpd, so that the no go theorem in \cite{roger}
may hopefully be evaded. An number of attempts at this problem have been 
made \cite{Takhtajan:1994vr,Chatterjee:1996ct,Hoppe:1997xp,Awata:2001dz}, 
but those
constructions, with the possible exception of the one based on Zariski 
quantisation \cite{Dito:1997xr}, have to drop some of 
the requirements imposed on the Nambu bracket, and seem rather ad hoc. 
It is a challenge to derive them from a proper quantisation of 
the topological open membrane.

\section{Clebsch parameterisation and Chern-Simons theory.}
In the previous section, we have argued that a divergenceless field
in 3 dimensions can be parameterized either by a gauge field,
$\xi=*dA$, or by a pair of scalar fields up to canonical transformation,
$\xi=*dH\wedge dK$. This implies that $A-H dK$ is closed, 
so that any gauge field $A$ can be represented locally in terms
of three scalar fields $(L,H,K)$ as $A=dL+H dK$. This is a familiar
representation in incompressible hydrodynamics, known as the 
Clebsch parameterisation of the velocity field $\xi$
(see e.g. \cite{Jackiw:2000mm} for a discussion of this representation). 
The Chern-Simons invariant $I=\int A \wedge dA$ of the 
Abelian gauge field $A$ is 
known as the total helicity, or asymptotic Hopf invariant
\cite{arnold}, and is
conserved in a perfect fluid. Expressed in terms of the Clebsch
parameterisation, it becomes $I=\int dL \wedge dH \wedge dK$,
our familiar membrane action again ! This seems to indicate that
the quantum membrane is equivalent to Abelian Chern-Simons on
a manifold with a boundary, or yet equivalently a $U(1)$ WZW model,
on the boundary, i.e. a free chiral scalar field in two 
dimensions~\footnote{A somewhat similar equivalence to $SU(2)$ WZW model
with the topological term only has been produced in
\cite{Rudychev:2001is}.}.
Unfortunately, the map $(L,H,K) \leftrightarrow A$ is very singular,
the $SO(3)$ symmetry is non-linearly realized in the gauge field
parameterization and the Jacobian it induces may have a 
non-trivial effect. On the
other hand, the Chern-Simons theory may be a sensible definition
of an otherwise ill-defined topological quantum membrane theory.

\section{Vpd and Nambu dynamics in higher dimensions.}
Our discussion of volume-preserving diffeomorphisms and Nambu dynamics
has so far been mostly restricted to the three-dimensional case (a).
However the topological membrane action \eqref{topact} in principle
makes sense in any dimension, and it is useful to recall some features
of three-form preserving diffeomorphisms in dimension $D>3$. 
Firstly, note that
a closed three-form $H$ has $(D-1)(D-2)/2$ degrees of freedom;
in contrast to the symplectic case, for $D>4$ it 
can therefore not be brought into a constant form by a 
diffeomorphism (however a self-dual three-form in $D=5+1$ has only 5
degrees of freedom so can be made constant by a diffeomorphism).
Second, assuming $H$ to be constant, for $D>5$ it still cannot be put 
into a standard form, but still carries moduli under the group of linear
diffeomorphisms $Gl(D,R)$ (for $D\leq 5$, it can be dualized into
a form of lower degree and a standard form exists). Third, for a 
given constant three-form $H_{ijk}$, three-form--preserving diffeomorphisms
$\xi^i=\zeta^i e^{ipx}$ have to satisfy the relation $H_{i[jk} p_{l]} 
\zeta^i=0$ (i) for all values of the indices $(jkl)$. In general, such a 
polarization $\zeta$ only exists for particular momenta $p$ 
(except again in $D \leq 5$, where $H$ can be dualized into a form
of lower degree). For special choices of $H$
however, satisfying the quadratic equation $H_{ij[k}H_{lm]p}=0$ (ii)
resulting from the elimination of $\zeta^i$ for all $p^i$ in (i), 
there is no constraint on $p$ anymore, and one
can have arbitrary momenta. This is in particular the case when $H$ 
degenerates to a single monomial such as in case (a) above. This
is also true in case (b), where any diffeomorphism $f(x_i)(\p_0+\p_1)$
preserves the form $H=(dx^0-dx^1)(dx^{23}+dx^{45})$ (of course, any
area-preserving diffeomorphisms in the (2345) would also preserve $H$). 
Conversely, we could try to define a Nambu bracket in higher dimension
as $\{H,K,F\}=\theta^{ijk}\p_i F\p_j K\p_k F$ where $\theta^{ijk}
\p_i\wedge\p_j\wedge\p_k$
is a three-vector. This definition automatically satisfies the
antisymmetry and derivation property. However the Fundamental Identity 
\eqref{fi} is only satisfied under special conditions, which in the
case of constant $\theta$ reduce to the algebraic condition
$\theta^{ij[k} \theta^{lmn]}= \theta^{kj[i} \theta^{lmn]}$ (iii) 
\cite{Takhtajan:1994vr}. This includes the degenerate case
(a) $\theta=\p_{123}$ but excludes non decomposable tensors
such as $\theta=\p_{123}+\p_{456}+\dots$ (in other words, Nambu dynamics 
lack extensivity). In particular, it excludes the lightlike
case (b) of interest for the non-commutative five-brane.
It would be interesting to see if the conditions (ii) or (iii) 
could be deduced from the quantum consistency of the membrane 
action \eqref{topact}, and if they bear any relation to the non-linear 
self-duality of the 3-form field strength $H$ on the five-brane
worldvolume.

\section{Vpd and non-Abelian two-form dynamics. \label{na2f}}
Despite providing some geometrical insight, the observations we have
presented so far have produced few results on the problem of the
quantum topological membrane. In the last part of this note,
we will propose a non-Abelian deformation of the dynamics of a 
two-form, based on the ideas developped before. Our approach is
a simple generalisation of an argument by Susskind and Bahcall
in the context of the Quantum Hall effect, relating the
dynamics of a perfect fluid in two dimensions to a gauge theory
for the group of area-preserving diffeomorphisms \cite{Bahcall:1991an}. 
Following on their steps, we consider a perfect fluid in 3+1 dimensions 
in comoving (Lagrange) coordinates,
\begin{equation}
{\cal L} = \int d^3x~ d\tau ~\left[ \frac{1}{2} 
\left( \frac{dy^i}{dt} \right)^2 - V\left(
\left|\det\frac{\partial x}{\partial y} \right| \right) \right]
\end{equation}
where $y^i(x,t)$ is the position of the fluid particle labelled by its
position at $x^i$ at $t=0$, where we assume a constant density 
normalized to 1 and also set the mass to 1. 
The potential $V$ describes short-range forces,
and is assumed to depend on the local density 
$\rho=|\det(\p x/\p y)|$ at time $t$ , with a stable minimum at $\rho=1$.
By construction, this Lagrangian is invariant under volume
preserving diffeomorphisms of the labelling coordinate $x^i$.
Now we consider small perturbations around equilibrium at $\rho=1$,
i.e. sound waves propagating in this perfect fluid. We parameterize
the fluctuations by a two-form, $y^i(x,t)=x^i+\epsilon^{ijk} b_{jk}(x,t)$.
Under a volume-preserving infinitesimal diffeomorphism $\xi$,
the field $y^i$ changes by $\xi^j(x,t)\p_j y^i$. Parameterizing
$\xi$ by a one-form $\xi^i=\epsilon^{ijk}\p_i a_j$, 
we find that the two-form $b$
transforms as
\begin{equation}
\label{dba}
\delta_\xi b_{ij} = \p_i a_j - \p_j a_i + \{a, b_{ij} \}
\end{equation}
where we employed the bracket $\{a,f\}:=\epsilon^{klm}\p_l a_m \p_k f$.
Equivalently, we may have parameterized $\xi$ by a pair of scalar
fields $\xi^i=\epsilon^{ijk}\p_j H \p_k K$, and expressed the 
variation of the two-form $b$ as 
\begin{equation}
\label{dbhk}
\delta_\xi b_{ij} = \p_i H \p_j K - \p_j H \p_i K + \{H,K,b_{ij}\}
\end{equation}
where the last term is the Nambu bracket \eqref{nambubra}. At leading order
in $b$, we find the usual $\delta b=da$ gauge variation
of an Abelian two-form, however the transformation receives a
correction linear in $b$, analogous to the Yang-Mills
variation for non-Abelian or non-commutative one-form gauge fields.
A gauge invariant field strength can be readily constructed by
considering the Jacobian $h_{123}:=\p y/\p x-1=1/\rho-1$, by construction
covariant under vpd. In terms of the two-form $b$, this is
\begin{eqnarray}
\label{hinv}
h_{123}&=&\left| \begin{matrix}
1+\p_1 b_{23} & \p_2 b_{23} & \p_3 b_{23} \\
\p_1 b_{31} & 1+ \p_2 b_{31} & \p_3 b_{31} \\
\p_1 b_{12} &  \p_2 b_{12} & 1+\p_3 b_{12} 
\end{matrix} \right| -1 \nn\\
&=&[\p_1 b_{23}+\mbox{circ}]+[\{b_{12},b_{13}\}_{23}+\mbox{circ}] \nn\\
&&\qquad\qquad\qquad\qquad +\{b_{12},b_{23},b_{31}\}
\end{eqnarray}
The leading term is indeed the usual exterior derivative $h=db+\dots$,
but there are now up to cubic terms in $b$, which ensure the gauge
covariance $\delta_\xi h_{ijk} = \{ a, h_{ijk} \} = \{H,K,h_{ijk}\}$
under vpd. In the above expression, $\{f,g\}_{ij}$
denotes the Poisson bracket in the plane $(ij)$, i.e.
$\{f,g\}_{ij}=\p_i f \p_j g - \p_j f \p_i g$.

Expanding the potential $V(\rho)$ to second order around its minimum at
$\rho=1$, we obtain for the Lagrangian for the sound waves described by the
two-form $b$,
\begin{equation}
\label{nalag}
{\cal L}=\int d^3 x~ d\tau~ \frac{1}{2\cdot 2!} \left( \p_\tau b_{ij} \right)^2
- \frac{1}{2\cdot 3!} h_{ijk}^2\ .
\end{equation}
In addition, one should impose the Gauss constraint restricting the
system to the gauge invariant sector. The conserved charge resulting
from invariance under vpd reads
\begin{equation}
Q=\int d^3 x \left[ \p_i a_j - \p_j a_i + \epsilon^{klm}\p_l a_m \p_k b_{ij} 
\right] \p_t b_{ij} 
\end{equation}
which vanishes for all gauge parameters $a_i$ under the Gauss constraint
\begin{equation}
\label{gauss}
G_i=2\p_j \p_\tau b_{ij} - \epsilon^{ijk} \p_j \p_\tau b_{lm} \p_k b_{lm} = 0
\end{equation}
This constraint can be enforced by means of a Lagrange multiplier 
$b_{0i}$, so that the Lagrangian takes the apparently Lorentz covariant form
${\cal L}=\int d^3x ~d\tau~ h_{\mu\nu\rho}^2$ with 
\begin{equation}
h_{0ij}=\p_0 b_{ij}+\p_j b_{0i}+\p_i b_{j0}+ \epsilon^{klm} \p_k b_{0l} 
\p_m b_{ij}
\end{equation}
Unfortunately, one can show that there is no choice of the gauge
variation $\delta_\xi b_{0i}$ that renders $h_{0ij}$ gauge covariant,
so that the free Lagrangian needs to be supplemented by a possibly
infinite sum of terms dependent on $b_{0i}$. In the temporal gauge
$b_{0i}=0$ however, the theory based on the action \eqref{nalag} 
supplemented by the Gauss constraint \eqref{gauss} is
perfectly consistent, and yields a non-Abelian deformation of
the dynamics of a two-form in 3+1 dimensions, based on 
the group of volume 
preserving diffeomorphisms~\footnote{It should be noted that this deformation
is different from the Freedman-Townsend vertex $\Tr(b\wedge [ *db, *db])$,
which still yields an Abelian gauge symmetry on-shell \cite{Freedman:us}.}. 
It would be
interesting to see if the algebraic structure of \eqref{dba},
\eqref{dbhk}, \eqref{hinv} 
can be abstracted, and the group of vpd replaced by other groups such as 
finite Lie groups. It is also important to generalize it to 5+1
dimensions, if one is to make contact with the five-brane.
As a possible step in this direction, and prompted by
the hydrodynamical picture advocated in \cite{Gibbons:2000ck}, one may 
consider a fluid of strings rather than particles: 
the same formulae as above still hold,
but for adding a fifth worldsheet coordinate $\sigma$, the
missing sixth coordinate being presumably due to
the decoupling of the $X^0-X^1$ coordinate in \eqref{topact}.
The effect of a background three-form would then be to add a
coupling $\int d^3x~d\tau~d\sigma~B_{12}\p_\sigma B_{23} \p_\tau B_{31}$
analogous to the Chern-Simons term in the Quantum Hall effect. 
It will be exciting to see whether M-theory or one of its avatars,
after having tamed the quantum Hall effect, can also master the
quantum vortex in superfluid He II.

\begin{acknowledgments}
{\it Acknowledgments.} 
These observations grew out of initial discussions with S. Minwalla 
and R. Gopakumar whom I gratefully acknowledge but who should not be 
blamed on any misconception herein. They were presented 
in May 2001 at the ``Avatars of M-theory'' program at ITP, 
Santa Barbara, whom I wish to thank for the kind hospitality. 
I also benefited from discussions with D. Berman, C. Hofman, 
D. Minic, Y. Matsuo, M. van Raamsdonk, S. da Silva and A. Strominger, and
I am grateful to C. Roger for communicating his results.
\end{acknowledgments}


\begin{thebibliography}{cc}

\bibitem{Callan:1991ky}
C.~G.~Callan, J.~A.~Harvey and A.~Strominger,
Nucl.\ Phys.\ B {\bf 367}, 60 (1991).


\bibitem{Witten:1995ex}
E.~Witten,
Nucl.\ Phys.\ B {\bf 443}, 85 (1995)
[arXiv:hep-th/9503124];
A.~Strominger,
Phys.\ Lett.\ B {\bf 383}, 44 (1996)
[arXiv:hep-th/9512059].

\bibitem{Ezawa:1997xc}
K.~Ezawa, Y.~Matsuo and K.~Murakami,
Phys.\ Rev.\ D {\bf 57}, 5118 (1998)
[arXiv:hep-th/9707200];
P.~Brax and J.~Mourad,
Phys.\ Lett.\ B {\bf 416}, 295 (1998)
[arXiv:hep-th/9707246];
C.~S.~Chu and E.~Sezgin,
JHEP {\bf 9712}, 001 (1997)
[arXiv:hep-th/9710223].




\bibitem{Dasgupta:2002iy}
A.~Dasgupta, H.~Nicolai and J.~Plefka,
arXiv:hep-th/0201182.

\bibitem{Bekaert:1999dp}
X.~Bekaert, M.~Henneaux and A.~Sevrin,
Phys.\ Lett.\ B {\bf 468}, 228 (1999)
[arXiv:hep-th/9909094].

\bibitem{Klebanov:1996un}
I.~R.~Klebanov and A.~A.~Tseytlin,
Nucl.\ Phys.\ B {\bf 475}, 164 (1996)
[arXiv:hep-th/9604089];
J.~A.~Harvey, R.~Minasian and G.~W.~Moore,
JHEP {\bf 9809}, 004 (1998)
[arXiv:hep-th/9808060];
M.~Henningson and K.~Skenderis,
JHEP {\bf 9807}, 023 (1998)
[arXiv:hep-th/9806087];

\bibitem{Connes:1997cr}
A.~Connes, M.~R.~Douglas and A.~Schwarz,
JHEP {\bf 9802}, 003 (1998)
[arXiv:hep-th/9711162].

\bibitem{Seiberg:1999vs}
N.~Seiberg and E.~Witten,
JHEP {\bf 9909}, 032 (1999)
[arXiv:hep-th/9908142].

\bibitem{Aganagic:1997zq}
M.~Aganagic, J.~Park, C.~Popescu and J.~H.~Schwarz,
Nucl.\ Phys.\ B {\bf 496}, 191 (1997)
[arXiv:hep-th/9701166];
P.~S.~Howe, E.~Sezgin and P.~C.~West,
Phys.\ Lett.\ B {\bf 399}, 49 (1997)
[arXiv:hep-th/9702008].


\bibitem{Bergshoeff:1987cm}
E.~Bergshoeff, E.~Sezgin and P.~K.~Townsend,
Phys.\ Lett.\ B {\bf 189}, 75 (1987);
E.~Bergshoeff, E.~Sezgin and P.~K.~Townsend,
Annals Phys.\  {\bf 185}, 330 (1988).

\bibitem{Bergshoeff:2000jn}
E.~Bergshoeff, D.~S.~Berman, J.~P.~van der Schaar and P.~Sundell,
Nucl.\ Phys.\ B {\bf 590}, 173 (2000)
[hep-th/0005026];
E.~Bergshoeff, D.~S.~Berman, J.~P.~van der Schaar and P.~Sundell,
Phys.\ Lett.\ B {\bf 492}, 193 (2000)
[hep-th/0006112].

\bibitem{Kawamoto:2000zt}
S.~Kawamoto and N.~Sasakura,
JHEP {\bf 0007}, 014 (2000)
[arXiv:hep-th/0005123].


\bibitem{Park:2000au}
J.~Park,
hep-th/0012141;
C.~M.~Hofman and W.~K.~Ma,
hep-th/0102201.

\bibitem{Matsuo:2001fh}
Y.~Matsuo and Y.~Shibusa,
JHEP{\bf 0102}, 006 (2001)
[hep-th/0010040].

\bibitem{Das:2001mg}
A.~K.~Das, J.~Maharana and A.~Melikyan,
JHEP {\bf 0104}, 016 (2001)
[arXiv:hep-th/0103229].

\bibitem{Gopakumar:2000ep}
R.~Gopakumar, S.~Minwalla, N.~Seiberg and A.~Strominger,
JHEP {\bf 0008}, 008 (2000)
[arXiv:hep-th/0006062].

\bibitem{Aharony:1997an}
O.~Aharony, M.~Berkooz and N.~Seiberg,
Adv.\ Theor.\ Math.\ Phys.\  {\bf 2}, 119 (1998)
[arXiv:hep-th/9712117];
O.~Aharony, J.~Gomis and T.~Mehen,
JHEP {\bf 0009}, 023 (2000)
[arXiv:hep-th/0006236];
M.~Berkooz,
arXiv:hep-th/0010158.


\bibitem{Lund:1976ze}
F.~Lund and T.~Regge,
Phys.\ Rev.\ D {\bf 14}, 1524 (1976).

\bibitem{kazhdan}
P.~Etingof, D.~Kazhdan, A.~Polishchuk, math.AG/ 0003009.

\bibitem{waldron}
D.~Kazhdan, B.~Pioline and A.~Waldron,
to appear in {\it Comm. Math. Phys.} [arXiv:hep-th/0107222];
B.~Pioline, H.~Nicolai, J.~Plefka and A.~Waldron,
JHEP {\bf 0103}, 036 (2001)
[arXiv:hep-th/0102123].




\bibitem{Rasetti:hk}
M.~Rasetti and T.~Regge, Physica {\bf 80A} (1975) 217; also
in proceedings of {\it ``Group Theoretical 
Methods In Physics''}, Trieste 1983, 311.

\bibitem{zee}
A.~Zee,
Nucl.\ Phys.\ B {\bf 421}, 111 (1994).

\bibitem{roger}
P.~Lecomte and C. Roger, 
J. Diff. Geom. {\bf 44} (1996) 529.

\bibitem{Nambu:1973qe}
Y.~Nambu,
Phys.\ Rev.\ D {\bf 7}, 2405 (1973).

\bibitem{Takhtajan:1994vr}
L.~Takhtajan,
Commun.\ Math.\ Phys.\ {\bf 160}, 295 (1994)
[hep-th/9301111].

\bibitem{Nambu:1980kz}
Y.~Nambu,
Phys.\ Lett.\ B {\bf 92}, 327 (1980).



\bibitem{Chatterjee:1996ct}
R.~Chatterjee and L.~Takhtajan,
Lett.\ Math.\ Phys.\ {\bf 37}, 475 (1996)
[hep-th/9507125].

\bibitem{Hoppe:1997xp}
J.~Hoppe,
Helv.\ Phys.\ Acta{\bf 70}, 302 (1997)
[hep-th/9602020].

\bibitem{Awata:2001dz}
H.~Awata, M.~Li, D.~Minic and T.~Yoneya,
JHEP{\bf 0102}, 013 (2001)
[hep-th/9906248].

\bibitem{Dito:1997xr}
G.~Dito, M.~Flato, D.~Sternheimer and L.~Takhtajan,
Commun.\ Math.\ Phys.\ {\bf 183}, 1 (1997)
[hep-th/9602016].

\bibitem{Jackiw:2000mm}
R.~Jackiw,
physics/0010042.

\bibitem{arnold}
V.I.~ Arnold, Sel. Math. Sov. {\bf 5} (1986) 327.

\bibitem{Rudychev:2001is}
I.~Rudychev,
JHEP {\bf 0104}, 015 (2001)
[arXiv:hep-th/0101039].

\bibitem{Bahcall:1991an}
S.~Bahcall and L.~Susskind,
Int.\ J.\ Mod.\ Phys.\ B {\bf 5}, 2735 (1991);
L.~Susskind,
hep-th/0101029.

\bibitem{Freedman:us}
D.~Z.~Freedman and P.~K.~Townsend,
Nucl.\ Phys.\ B {\bf 177}, 282 (1981).


\bibitem{Gibbons:2000ck}
G.~W.~Gibbons and P.~C.~West,
hep-th/0011149.



\end{thebibliography}
\end{document}